\documentclass[a4paper, 10pt, twocolumn]{article}

\usepackage{fancyhdr}

\newif\ifpdf 
\ifx\pdfoutput\undefined 
\else 
\pdfoutput=1 
\pdftrue 
\fi

\ifpdf
\usepackage[pdftex,bookmarks=true,bookmarksopen=true,bookmarksnumbered=true]{hyperref}
\hypersetup{%
pdftitle = {An Exact Formula to Describe the Amplification Process in a Photomultiplier Tube},
pdfauthor = {Jonas Rademacker},
pdfkeywords = {PMT, photo multiplier tube, dynode, Poisson, exact formula, fit PMT spectra, dynode chain, MaPMT, LHCb, RICH}
}
\usepackage[pdftex]{graphicx}
\else
\usepackage[dvips,bookmarks=true,bookmarksopen=true,bookmarksnumbered=true]{hyperref}
\usepackage[dvips]{graphicx}
\fi

\usepackage{rotating}

\fontsize{12}{24}
\newcommand{\C}{Cherenkov}

\newcommand{\fig}{.}

\newcommand{\labelpawx}[6]{heinz}

\newcommand{\labelpaw}[6]{
\parbox[t]{1ex}{\mbox{}\\[-\baselineskip] \rotatebox{90}{#5\hspace{1em}}}\parbox
[t]{#2}{
\mbox{}\\[-\baselineskip]
\parbox[b]{#2}{\includegraphics[width = #3 ]{#1} }\vspace{#6}
\\[-\baselineskip]\mbox{}\hfill #4 \hspace{1em}}
}

\newcommand{\labeltrimmedpaw}[6]{
\parbox[t]{1ex}{\mbox{}\\[-\baselineskip]
\rotatebox{90}{#5\hspace{1em}}}\parbox
[t]{#2}{
\mbox{}\\[-\baselineskip]
\parbox[b]{#2}{\mbox{}\vspace{#6}\\\mbox{}\hspace{0.7em}
\includegraphics[width = #3 ]{#1} }\vspace{#6}
\\[-\baselineskip]\mbox{}\hfill #4 \hspace{1em}}
}

\begin{document}
\onecolumn

\title{An Exact Formula to Describe the Amplification Process in a
Photomultiplier Tube}
\author{Jonas Rademacker
\\ \textit{University of Oxford}
}
\date{}
%
\maketitle
\fancyfoot{}
  \renewcommand\headrulewidth{0pt}
  \renewcommand\footrulewidth{0pt}
\setcounter{page}{0}
\thispagestyle{fancy}
\vfill
\begin{abstract}
An analytical function is derived that exactly describes the
amplification process due to a series of discrete, Poisson-like
amplifications like those in a photo multiplier tube (PMT). A
numerical recipe is provided that implements this function as a
computer program. It is shown how the program can be used as the
core-element of a faster, simplified routine to fit PMT spectra with
high efficiency.  The functionality of the method is demonstrated by
fitting both, Monte Carlo generated and measured PMT spectra.
\end{abstract}
\twocolumn
\section{Introduction}
In September 1999, the LHCb RICH group tested Hamamatsu's 64-Multi-anode
Photo Multiplier Tubes as a possible photodetector choice for the
LHCb RICH detector \cite{lhcb:rich.tdr}.  During the data analysis, the
need for an accurate model of the output of a PMT arose that could be
fitted to the measured pulse height spectra, mainly in order to have a
precise estimate of the signal lost below the threshold cut. In order
to perform a fit to the spectra, an analytical function is needed that
can be calculated reasonably quickly by a computer.

Such a function is derived here. First (section \ref{sec:derive_p}),
an analytical expression is derived that describes the output of a
PMT. The starting assumption is that the number of photoelectrons per
event, as well as the number of secondary electrons caused by each
primary electron at each stage of the dynode chain, are well described
by Poisson distributions. Furthermore it is shown how this expression
can be adapted to avoid some of the numerical problems arising in the
original expression, so that it can be calculated by a computer. A
complete numerical recipe is given and a FORTRAN implementation of the
program is listed in appendix
\ref{app:listing}. This expression can of course be used to calculate
any ``snowball'' like effect described by a series of Poisson
distributions.

In section \ref{sec:fitfunction} it is described how the exact
expression derived in the first part can be used as the central
element of a faster, approximate function, and how the number of
parameters can be reduced making reasonable assumptions, so that
fitting a large number of spectra in a finite time becomes
feasible. This is then adapted to describe the digitised output of
laboratory read-out electronics, rather than the number of electrons
at the end of a dynode chain.

This approximate function is used in section \ref{sec:examplefits} of the
paper to fit Monte Carlo generated spectra as well as real data,
demonstrating the validity of the method.
\section{An Analytical Function}
\label{sec:derive_p}
\subsection{The Electron Probability Distribution} 
\label{sec:derive_basic}
In the following, an expression for the number of photoelectrons at
the end of a dynode chain of a PMT is derived. The number of incident
photons, and hence of photoelectrons produced in the cathode, is
assumed to follow a Poisson distribution. This is appropriate for the
testbeam data where PMTs were used to detect \C\ photons generated by
a particle traversing a dielectric.  With a mean number of
photoelectrons produced in the cathode of $\lambda_1$, the probability
to find $k_1$ photoelectrons arriving at the first dynode is:
\begin{equation}
\label{eqn:poisson_k1}
P(k_1)=e^{-\lambda_1} \frac{\lambda_1^{k_1}}{k_1!}.
\end{equation}
The probability to find $k_2$ electrons after the first dynode is the
sum over all values for $k_1$ of the probabilities $P(k_1)$, each
multiplied by the probability that the dynode returns $k_2$ electrons
given that $k_1$ arrive:
\begin{equation}
\label{eqn:k2_start}%
P(k_2) =%
\sum_{k_1 = 0}^{\infty}%
P(k_1)\cdot P(k_2|k_1).
\end{equation}\\
Each of the $k_1$ electrons produces a Poisson--distributed response
from the dynode with mean $\lambda_2$ where $\lambda_2$ is the gain at
the $1^{\mathrm{st}}$ dynode; all $k_1$ electrons together produce a
response distributed according to the convolution of $k_1$ Poisson
distributions, each with mean $\lambda_2$. This results in a single
Poisson distribution with mean $\lambda_2 \cdot k_1$:
\begin{equation}
P(k_2|k_1)=e^{-\lambda_2 k_1}\frac{\left(\lambda_2 k_1\right)^{k_2}}{k_2!}.
\end{equation}
\\
Hence the probability to find $k_2$ electrons after the first dynode
is given by:
\begin{equation}
\label{eqn:k2}%
P(k_2) =%
\sum_{k_1 = 0}^{\infty}%
P(k_1) \cdot e^{-\lambda_2 k_1}\frac{\left(\lambda_2 k_1\right)^{k_2}}{k_2!}.
\end{equation}
\\
Inserting the right--hand side of equation \ref{eqn:poisson_k1} for
$P(k_1)$ yields, after manipulation:
\begin{equation}
\label{eqn:k2_solved}
P(k_2) = e^{-\lambda_1} \frac{\lambda_2^{k_2}}{k_2!}
\sum_{k_1 = 0}^{\infty}
\frac{(\lambda_1 e^{-\lambda_2})^{k_1}}{k_1!} k_1^{k_2}.
\end{equation}
\\
Generalising this for $n-1$ dynodes yields:\\
\parbox{\columnwidth}{\small
\begin{eqnarray}
\label{eqn:for_n_dynodes}
P(k_n) &=& e^{-\lambda_1} \frac{\lambda_n^{k_n}}{k_n!} \nonumber\\
   & & \sum_{k_1 = 0}^{\infty} \sum_{k_2 = 0}^{\infty}
   \cdots \sum_{k_{n-1} = 0}^{\infty}  \nonumber\\
   & & \frac{(\lambda_1 e^{-\lambda_2})^{k_1}}{k_1!} 
   \frac{(\lambda_2 e^{-\lambda_3} k_1)^{k_2}}{k_2!} \nonumber\\
   & &
   \frac{(\lambda_3 e^{-\lambda_4} k_2)^{k_3}}{k_3!} \cdots \nonumber\\
   & & \frac{(\lambda_{n-2} e^{-\lambda_{n-1}}
   k_{n-3})^{k_{n-2}}}{k_{n-2}!}\nonumber\\ 
   & & \frac{(\lambda_{n-1} e^{-\lambda_{n}} k_{n-2})^{k_{n-1}}}{k_{n-1}!}
   k_{n-1}^{k_n} .
  \nonumber\\
  \mbox{}
\end{eqnarray}
}\\ Each term in equation \ref{eqn:for_n_dynodes} is of the form of an
exponential series, i.e. $\frac{x^k}{k!}$, except for the last term
with the summation parameter $k_{n-1}$, which appears as
$\frac{x^k}{k!} k^{k_n}$.  This term can be expressed in terms of the
$k_n$th derivative of $e^{y k_{n-1}}$ with respect to the new variable
$y$ at $y=0$:\\
\begin{equation}
k_{n-1}^{k_n}= \frac{\mathsf{d}^{k_n}}{\mathsf{d} y^{k_n}} e^{y
k_{n-1}} \Big|_{y=0}
.
\end{equation}
\\
Now the last term in equation \ref{eqn:for_n_dynodes} can be written
as\\
\parbox{\columnwidth}{\small
\begin{eqnarray}
\label{eqn:last_term}
&& \frac{(\lambda_{n-1} e^{-\lambda_{n}} k_{n-2})^{k_{n-1}}}{k_{n-1}!}
   k_{n-1}^{k_n}\nonumber\\
&=&
\frac{\mathsf{d}^{k_n}}{\mathsf{d} y^{k_n}} \frac{(\lambda_{n-1} e^{-\lambda_{n}}
k_{n-2} e^{y})^{k_{n-1}}}{k_{n-1}!} \bigg|_{y=0} 
.
\end{eqnarray}
}\\
Using \ref{eqn:last_term}, equation \ref{eqn:for_n_dynodes} can be
re--written as: 
\parbox{\columnwidth}{\small
\begin{eqnarray}
\label{eqn:for_n_dynodes_rewritten}
P(k_n) &=& e^{-\lambda_1} \frac{\lambda_n^{k_n}}{k_n!}
   \;\;\;\;\;
   \frac{\mathsf{d}^{k_n}}{\mathsf{d} y^{k_n}}
   \nonumber \\
   & & \sum_{k_1 = 0}^{\infty} \sum_{k_2 = 0}^{\infty}
   \cdots \sum_{k_{n-1} = 0}^{\infty} \nonumber \\
   & & \frac{(\lambda_1 e^{-\lambda_2})^{k_1}}{k_1!} 
   \frac{(\lambda_2 e^{-\lambda_3} k_1)^{k_2}}{k_2!} \nonumber \\
   & &
   \frac{(\lambda_3 e^{-\lambda_4} k_2)^{k_3}}{k_3!} \cdots \nonumber \\
   & & \frac{(\lambda_{n-2} e^{-\lambda_{n-1}}
   k_{n-3})^{k_{n-2}}}{k_{n-2}!} \nonumber \\
   & &
   \frac{(\lambda_{n-1} e^{-\lambda_{n}} k_{n-2}
   e^{y})^{k_{n-1}}}{k_{n-1}!}\bigg|_{y=0} . \nonumber \\
   \mbox{}
\end{eqnarray}
}\\
Now each summation can be carried out in turn, starting with that over
$k_{n-1}$: 
{\small
\begin{eqnarray}
\lefteqn{\sum_{k_{n-1} = 0}^{\infty}
   \frac{(\lambda_{n-1} e^{-\lambda_{n}} k_{n-2}
   e^{y})^{k_{n-1}}}{k_{n-1}!}}\nonumber \\
&=& \exp(\lambda_{n-1} e^{-\lambda_{n}} k_{n-2}
e^{y}) \nonumber \\
&=& (\exp(\lambda_{n-1} e^{-\lambda_{n}}
e^{y}))^{k_{n-2}},
\nonumber\\
\mbox{}
\end{eqnarray}
}
then over $k_{n-2}$:
{\small
\begin{eqnarray}
\lefteqn{\sum_{k_{n-2} = 0}^{\infty} \Big(
\frac{(\lambda_{n-2} e^{-\lambda_{n-1}}
   k_{n-3})^{k_{n-2}}}{k_{n-2}!}\cdot}\nonumber\\ 
&&  (e^{(\lambda_{n-1} e^{-\lambda_{n}}
    e^{y})})^{k_{n-2}} \Big)
\nonumber \\
&=&\exp(\lambda_{n-2} e^{-\lambda_{n-1}}
   k_{n-3}\cdot e^{(\lambda_{n-1} e^{-\lambda_{n}}
  e^{y})}) 
\nonumber \\
&=&(\exp(\lambda_{n-2} e^{-\lambda_{n-1}}
  \cdot e^{(\lambda_{n-1} e^{-\lambda_{n}}
  e^{y})}))^{k_{n-3}} 
\nonumber\\
  \mbox{}
\end{eqnarray}
}
\\and so on.
After performing all these summations, the probability of finding
$k_n$ electrons after $n-1$ dynodes, with gains $\lambda_2,\ldots,
\lambda_n$, starting off with an average of $\lambda_1$ photo
electrons arriving at the first dynode, is given by:
\vspace{0.5ex}\\
\fbox{\parbox{0.95\columnwidth}{
\begin{eqnarray}
\label{eqn:analyticsolution}
P(k_n) &=& e^{-\lambda_1} \frac{\lambda_n^{k_n}}{k_n!}
  \;\;\;\;\;
  \frac{\mathsf{d}^{k_n}}{\mathsf{d}y^{k_n}}\nonumber\\
  & & \exp(x_1 \exp( x_2 \exp( x_3 \cdots \nonumber\\
  & & \exp(x_{n-1} 
  \exp(y) ) \cdots ))) \bigg|_{y=0} \nonumber \\
  \mathrm{with}\; x_i &\equiv& \lambda_i e^{-\lambda_{i+1}} .\nonumber\\
  \mbox{}
\end{eqnarray}
}}
\subsection{Calculating $P(k_n)$}
%
In order to calculate $P(k_n)$ it is useful to make the following
definitions:\\
\parbox{\columnwidth}{
\begin{eqnarray}
\label{eqn:define_f}
f_1 & \equiv & e^{x_1 e^{x_2...e^{x_{n-1} e^{y}}}} \nonumber \\
f_2 & \equiv & e^{x_2 e^{x_3...e^{x_{n-1} e^{y}}}} \nonumber \\
f_3 & \equiv & e^{x_3 e^{x_4...e^{x_{n-1} e^{y}}}} \nonumber \\
    &\vdots& \\
f_{n-1} & \equiv & e^{x_{n-1} e^{y}} \nonumber \\
f_{n} & \equiv & e^{y} . \nonumber
\end{eqnarray}
}
Equation \ref{eqn:analyticsolution} can now be written as:
\\
\parbox{\columnwidth}{
\begin{equation}
\label{eqn:solution_with_f}
P\left(k_n\right) = e^{-\lambda_1} \frac{\lambda_n^{k_n}}{k_n!}
f_1^{(k_n)}(y)\bigg|_{y=0} ,
\end{equation}
}
where $ f_1^{(k_n)}$ is the $k_{n}th$ derivative of $f_1$ with respect
to $y$.  With the above definitions, the first derivatives of the
functions $f_i$ are given by:\\
\parbox{\columnwidth}{
\begin{equation}
\begin{array}{l r *{5}c}
f_1^{\prime} & = & f_1 x_1 & f_2 x_2 &\cdots &f_{n-1} x_{n-1} & f_n \\
f_2^{\prime} & = &         & f_2 x_2 &\cdots &f_{n-1} x_{n-1} & f_n \\
f_3^{\prime} & = &         &         &\cdots &f_{n-1} x_{n-1} & f_n \\
             &\vdots &     &         &       &                &     \\
f_n^{\prime} & = &         &         &       &                & f_n .
\end{array}
\end{equation}
}
This gives a  recursive formula for the first derivative of $f_{i}$:
\begin{eqnarray}
f_{i}^{\prime} & = & f_{i} x_{i} f_{i+1}^{\prime} \;\;\;\; i<n\nonumber\\ 
f_n^{\prime} & = & f_n ,
\end{eqnarray}
which in turn gives a recursive formula for the $m$th derivative:
\begin{equation}
\label{eqn:recursive}
 f_i^{(m)} =\sum_{j=0}^{m-1}{m-1\choose
 j}f_i^{(j)}x_{i}f_{i+1}^{(m-j)} ,
\end{equation}\\
\begin{displaymath}
\mathrm{with}\; f_n^{(j)} = f_n \;\; \forall j \in \mathsf{I\!N}.
\end{displaymath}\\
With this expression, equation \ref{eqn:solution_with_f} can finally
be calculated, by starting with $f_n(0)=1$ and calculating $f_i^{(m)}$
subsequently for all values $i=n, n-1, \ldots, 1$ and all values
$m=0,1, \ldots, k_n$.
\subsection{Numerical Difficulties}
While the previous section gives a valid algorithm on how to calculate
$P(k_n)$ using equation \ref{eqn:solution_with_f} and the recursive
formula \ref{eqn:recursive}, it turns out that the finite precision of
a normal computer will only allow calculations to be performed for
rather small values of $k_n$ before some numbers become either too
large or too small to be stored straightforwardly in the computer
memory. This problem is addressed in the following discussion.

\paragraph{The factor ${\bf \frac{\lambda_n^{k_n}}{k_n!}}$}
For any reasonably large number of dynodes, where the mean number of
electrons coming off the last dynode, and therefore the interesting
values for $k_n$, is typically in the
thousands or even millions, \(e^{-\lambda_1}
\frac{\lambda_n^{k_n}}{k_n!} \) quickly becomes very small, while
\(f_1^{(k_n)}(y)\bigg|_{y=0}\) grows to extremely large values. In
order to calculate $P(k_n)$ for such values of $k_n$, it is
necessary to absorb the small factor \( \frac{\lambda_n^{k_n}}{k_n!}\)
into the $f_i^{(m)}$. This can be done by replacing
$y$ in equation \ref{eqn:solution_with_f} with $py$ and introducing a
compensating factor $\left(\frac{1}{p}\right)^{k_n}$:
\begin{equation}
\label{eqn:compensate}
P\left(k_n\right) = e^{-\lambda_1} 
\frac{\lambda_n^{k_n}}{k_n!}
\left(\frac{1}{p}\right)^{k_n}
\frac{\mathsf{d}^{k_n}}{\mathsf{d}y^{k_n}} 
f_1(py)\bigg|_{y=0} .
\end{equation}
Choosing $p$ such that $p^{k_n}=\frac{\lambda_n^{k_n}}{k_n!}$ changes
equation \ref{eqn:solution_with_f} to
\begin{eqnarray}
\label{eqn:solution_with_p}
P\left(k_n\right) & =&  e^{-\lambda_1} 
\frac{\mathsf{d}^{k_n}}{\mathsf{d}y^{k_n}}
f_1(py)\bigg|_{y=0}
\nonumber\\
\mbox{with} & & p^{k_n}=\frac{\lambda_n^{k_n}}{k_n!} .
\end{eqnarray}
Defining
\begin{eqnarray}
\label{eqn:def_fstar}
f^{\star(m)}_{k_n,i} \equiv 
\frac{\mathsf{d}^{m}}{\mathsf{d}y^{m}} f_i(p_{k_n}y)\bigg|_{y=0} 
\\
\mbox{with}\;\;
p_{k_n}=\frac{\lambda_n}{\left(k_n!\right)^{\frac{1}{k_n}}}
\nonumber
\end{eqnarray}
gives
\begin{equation}
\label{eqn:solution_with_fstar}
P\left(k_n\right) =  e^{-\lambda_1} 
f^{\star(k_n)}_{k_n,1} .
\end{equation}
The recursive formula established for calculating $f_1^{(k_n)}$
remains essentially the same for $f^{\star(k_n)}_{k_n,1}$:
\begin{equation}
\label{eqn:recursive_star}
 f_{k_n,i}^{\star (k_n)} =\sum_{j=0}^{k_n-1}{k_n-1\choose
 j}f_{k_n,i}^{\star (j)}x_{i}f_{k_n,i+1}^{\star (k_n-j)}\\
\end{equation}\\
\begin{displaymath}
\mathrm{with}\; f_{k_n,n}^{\star (m)} = 
p_{k_n}^m \;\; 
\mathrm{and}\;\; p_{k_n}=
\frac{\lambda_n}{\left(k_n!\right)^{\frac{1}{k_n}}},
\end{displaymath}\\ 
with one additional complication. In the original algorithm, when
calculating $f_1^{k_n}$ using the recursive formula
\ref{eqn:recursive}, all values for $f_i^m$ with $m<k_n$ calculated in
the previous iterations\footnote{where $P\left(0\right),\ldots,
P\left(k_n-1\right)$ were calculated} could be used in the recursive
formula for the current iteration. Now, for calculating
$f_{k_n,1}^{\star (k_n)}$ \emph{all} values $f_{k_n,i}^{\star (m)}$
with $m < k_n$, $i\le n$ have to be re--calculated at each iteration,
because at each iteration the value for $p$ in equation
\ref{eqn:recursive_star} changes.
To calculate $f_{k_n,i}^{\star(k_n)}$, from equation
\ref{eqn:recursive_star}, the values for\\
\[
f_{k_n,i}^{\star(m)}, m<k_n
\]\\
are needed. These can be calculated using only the values for
$f_{k_n-1,i}^{\star(m)}$ which have been calculated one iteration
earlier:
\[
f_{k_n,i}^{\star (m)} = f_{k_n-1,i}^{\star (m)} 
\left(\frac{p_{k_n}}{p_{k_n-1}}\right)^{m}
\]
with\\
\begin{equation}
\label{eqn:pk_by_pk-1}
\left(\frac{p_{k_n}}{p_{k_n-1}}\right)^{k_n} 
=
\frac
{ \left(\left(k_n-1\right)!\right)^{1/(k_n-1)}}{k_n} ;
\end{equation}\\
so the values for $f_{k_n,i}^{\star(k_n)}$ need to be stored only for
one iteration.

\paragraph{The binomial factor}
When calculating $f_{k_n,i}^{\star (k_n)}$, using the recursive formula
\ref{eqn:recursive_star}, the factor ${k_n-1\choose j}$ in 
\[
f_{k_n,i}^{\star (k_n)} =\sum_{j=0}^{k_n-1}{k_n-1\choose
j}f_{k_n,i}^{\star (j)}x_{i}f_{k_n,i+1}^{\star (k_n-j)}
\]
can get very large for large values of $k_n$, while the corresponding
values for \(f_{k_n,i}^{\star (j)}x_{i}f_{k_n,i+1}^{\star (k_n-j)}\)
get very small. To avoid the associated numerical problems, one can
define the arrays $u_{k_n,i}^{(j)}$ and $v_{k_n,i}^{(j)}$ that
`absorb' the binomial factor, such that equation
\ref{eqn:recursive_star} becomes:\\
\begin{equation}
\label{eqn:recursive_uv}
f_{k_n,i}^{\star (k_n)}= \sum_{j=0}^{k_n-1} u_{k_n,i}^{(j)} x_{i}
v_{k_n,i+1}^{(k_n-j)} ,
\end{equation}
where
\begin{eqnarray}
\label{eqn:define_uv}
u_{k_n,i}^{(j)}= \sqrt{{k_n-1 \choose j}} f_{k_n,i}^{\star (j)}\nonumber\\
v_{k_n,i}^{(j)}= \sqrt{{k_n-1 \choose {j-1}}} f_{k_n,i}^{\star (j)} .
\end{eqnarray}
\subsection{Combining Results}
At each iteration $k_n$, before calculating $f_{k_n,i}^{\star (k_n)}$
using equation \ref{eqn:recursive_uv}, the values for
$u_{k_n,i}^{(j)}$ and $v_{k_n,i}^{(j)}$, $j<k_n$, are calculated from
their values in the previous iteration:
\begin{eqnarray}
\label{eqn:uvk_from_uvk-1}
u_{k_n,i}^{(j)} &=& \left(\frac{p_{k_n}}{p_{k_n-1}}\right)^{j} 
\sqrt{\frac{k_n-1}{k_n-1 - j}} u_{k_n-1,i}^{(j)} 
\nonumber\\
v_{k_n,i}^{(j)} &=& \left(\frac{p_{k_n}}{p_{k_n-1}}\right)^{j} 
\sqrt{\frac{k_n-1}{k_n - j}} v_{k_n-1,i}^{(j)}
\nonumber\\
j & < & k_n .
\end{eqnarray}
These results are then used to calculate:
\begin{equation}
\label{eqn:fstar_binom}
f_{k_n,i}^{\star (k_n)}=
\sum_{j=0}^{k_n-1} u_{k_n,i}^{(j)} x_{i} v_{k_n,i+1}^{(k_n-j)}
\end{equation}
and
\begin{equation}
u_{k_n,i}^{(k_n)}=v_{k_n,i}^{(k_n)}=f^{\star (k_n)}_{k_n,i} ,
\end{equation}
starting from
\begin{equation}
u_{k_n,n}^{(k_n)}=v_{k_n,n}^{(k_n)}=f_{k_n,n}^{\star (k_n)}=
\frac{\lambda_n^{k_n}}{k_n!}
\end{equation}
and
\begin{equation}
u_{0,i}^{(0)}=v_{0,i}^{(0)}=f_{0,i}^{\star (0)}=f_i ,
\end{equation}
where the $f_i$ are defined by equation \ref{eqn:define_f}.

\subsection{The Complete Numerical Recipe}
%
%
Using the above formulae, the problem of calculating the probability
distribution of finding $k_n$ electrons at the end of a PMT with $n-1$
dynodes can be solved by a computer. A FORTRAN implementation is listed
in appendix \ref{app:listing}.
The program takes as its input the array $\lambda[n]$, with dimension
n, which contains the average number of photo electrons arriving at
the first dynode $\lambda[1]$ and the gain at each of the $n-1$
dynodes, $\lambda[2], \ldots, \lambda[n]$. The program fills the
array $P[\mathsf{max}]$ with the probabilities $P[k]$ to find $k$
electrons at the end of the dynode chain for all values $k\le
\mathsf{max}$. The parameter $\mathsf{max}$ is also passed to the
program.

The values for $u_{k,i}^{(j)}, v_{k,i}^{(j)}$ needed in the recursive
formulae, are stored in two two-dimensional arrays, where one
dimension is taken by the index $i=1,\ldots,n$, and the other by the
index $j=0,\ldots,\mathsf{max}$. As the values for $u_{k,i}^{(j)},
v_{k,i}^{(j)}$ are needed only for one value of $k$ at a time, the
arrays do not need to be three-dimensional; the values for
$u_{k,i}^{(j)}, v_{k,i}^{(j)}$ needed at the iteration calculating
$P[k]$ replace those from the previous iteration,
$u_{k-1,i}^{(j)},v_{k-1,i}^{(j)}$.

The steps to calculate $P[k], k=0,\ldots,\mathsf{max}$ are:

\begin{itemize}
\item[{\bf 1}] Initialise program, test whether input is sensible, for
example if the overall gain is larger than 0. Calculate all values for
$\left(\frac{p_{j}}{p_{j-1}}\right)^{j}, j\le \mathsf{max}$ and store them in
an array $p_{\mathsf{frac}}[j],j=1,\ldots\mathsf{max}$ for later use.
\item[{\bf 2}] Start with calculating the probability to find zero
electrons: $k=0$
\item[{\bf 3}] Calculate $u_{0,i}^{(0)}=v_{0,i}^{(0)}=f_i$ for $i=n, n-1,
\ldots, 1$, as defined by equation \ref{eqn:define_f}
\item[{\bf 4}] Store the result in the array: $P[0]=e^{-\lambda_1}
u_{0,1}^{(0)}$
\item[{\bf 5}] Increment $k$ by 1. If $k>\mathsf{max}$, stop program.
\item[{\bf 6}] Calculate $u_{k,n}^{(k)}=v_{k,n}^{(k)}=\frac{\lambda_n^{k}}{k!}$
\item[{\bf 7}] Calculate $u_{k,i}^{(j)}, v_{k,i}^{(j)}$ for $j<k$ and
\mbox{$i=n,\ldots,1$} from $u_{k-1,i}^{(j)}, v_{k-1,i}^{(j)}$ according to
equation \ref{eqn:uvk_from_uvk-1}, using the values of
$p_{\mathsf{frac}}[k]$ calculated in step 1.
\item[{\bf 8}] Calculate $u_{k,i}^{(k)}=v_{k,i}^{(k)}$ for all values
of $i<n$ using the recursive formula \ref{eqn:fstar_binom}. Let the
outer loop go over the index $i$, starting with $i=n-1$ and
decrementing it by 1 until $i=1$, and the inner loop over the
summation index $j$, starting with $j=0$ and incrementing $j$ by $1$
until $j=k-1$.
\item[{\bf 9}] Store result: $P[k]=e^{-\lambda_1}u_{k,1}^{(k)}$
\item[{\bf 10}] Goto step 5
\end{itemize}

\section{Fitting PMT Spectra}
\label{sec:fitfunction}
\subsection{Increasing Speed by Approximating ${\bf P(k_n)}$}
When fitting PMT--pulse--height spectra, speed is a major problem. The
number of operations needed to calculate $P(k_n)$ using the recursive
formula in equation \ref{eqn:recursive}, is\\
\begin{equation}
N_{\mathsf steps}\approx \sum_{i=0}^{k_n} \sum_{j=0}^{i} nj \sim
k_n^3 ,
\end{equation}\\
which becomes prohibitive for a typical PMT with a gain of $\sim 10^5$
and higher. Therefore, for fitting the spectra, only the exact
distribution after the first $m$ dynodes is calculated and then scaled
by the gain of the remaining dynodes, $g_{\mathsf{left}}=
\left(g_{m+1} g_{m+2} \cdots g_{n-1}\right)$. When scaling the output
of the exact distribution calculated for the first $m$ dynodes,
$P_{\mathsf{exact}}(k_{m+1})$, to the final distribution, the result
is convoluted with a Gaussian of width $\sigma_{\mathsf{scale}}$,
taking into to account the additional spread in the distribution at
each remaining dynode:
\begin{equation}
\label{eqn:sigmascale}
\sigma_{\mathsf{scale}} =\sqrt{k_{m+1}}\sigma_0
\end{equation}
with:
{\small
\begin{equation}
\begin{array}{rl}
\sigma_0 &=
\left(g_{m+1} g_{m+2}
\cdots g_{n-1}\right)
\\
&
\cdot \left( \frac{1}{g_{m+1}} + \frac{1}{g_{m+1}g_{m+2}} + \cdots +
\frac{1}{g_{m+1}\cdots g_{n-1}}\right)^{\frac{1}{2}} .
\end{array}
\end{equation}
}
So the approximated function, $P_{\mathsf{\sim}}(k_n)$ is
\begin{equation}
\label{eqn:scale}
P_{\mathsf{\sim}}(k_n) = 
\sum_{j=0}^{\infty} \frac{1}{\sqrt{2\pi}\sqrt{j}\: \sigma_0}
e^{ 
\frac{\left(j\cdot g_{\mathsf{left}} - k_n \right)^2}{
2\left(\sqrt{j}\: \sigma_0\right)^2}} P(j) .
\end{equation}
In practice the sum only needs to be calculated for values of $j\cdot
g_{\mathsf{left}}$ that are a few $\sigma_{\mathsf{scale}}$ around
$k_n$.
\subsection{Reducing the Number of Parameters}
$P(k_n)$ depends on n parameters: the gain of each dynode and the
number of photoelectrons produced in the cathode. For the case of the
12--dynode PMT, there are 13 parameters. It is possible, however, to
reduce this number to two:
\begin{enumerate}
\item the mean number of photoelectrons produced in the
photo cathode
\item the gain at the first dynode.
\end{enumerate} 
Using\\ 
\begin{equation}
g \propto V^{\alpha},
\end{equation}
where V is the voltage difference over which the electron is
accelerated, the gain at the other dynodes can be calculated from the
gain at the first dynode. The parameter $\alpha$ has values typically
between $0.7$ and $0.8$ \cite{hamamatsu}; in the following,
$\alpha=0.75$ is used.
\subsection{Adapting the Function to Fit Measured Data}
 In practice, spectra are not measured in numbers of photoelectrons,
 but in ADC counts digitised by the readout electronics.  The function
 describing the spectra needs to relate the ADC counts,
 $k_{\mathsf{adc}}$, to the number of electrons at the end of the
 dynode chain, $k_n$. This requires two parameters: the offset, or
 pedestal mean, $p_0$, and the conversion factor, $c_n$ of $k_n$ to
 ADC counts. The resulting function is convoluted with a Gaussian of
 width $\sigma$ to take into account electronics noise:\\
\parbox{0.98\columnwidth}{\small
\begin{eqnarray}
\lefteqn{F_{\mathsf{cont}}(k_{\mathsf{adc}}) = }
\nonumber\\
&& \left(\frac{1}{\sqrt{2\pi}\sigma}
e^{\frac{k_{\mathsf{adc}}^2}{2\sigma^2}}\right) \ast
\left(P\left(\left(k_{\mathsf{adc}}-p_o\right)/c_n\right) \cdot
c_n\right), 
\nonumber \\
\mbox{} & &
\end{eqnarray}
}
\\
where $\ast$ is the convolution operator. $F_{\mathsf{cont}}$ treats
$k_{\mathsf{adc}}$ as a continuous variable, with a one--to--one
relation between $k_{\mathsf{adc}}$ and $k_n$; in fact the readout
electronics deliver only integer--value ADC counts, integrating over
the corresponding pulse heights. Thus the final function for describing
ADC spectra is:
\begin{equation}
F(k_{\mathsf{adc}})=
\int_{k_{\mathsf{adc}}-0.5}^{k_{\mathsf{adc}}+0.5}
F_{\mathsf{cont}}(k^{\prime}_{\mathsf{adc}})
\mathsf{d}\!k^{\prime}_{\mathsf{adc}} .
\end{equation}

\section{Example Fits}
\label{sec:examplefits}
The fits are performed as binned log--likelihood fits: for each
1--ADC--count wide bin $k_{\mathsf{adc}}$, containing $n_i$ events, the
binomial probability of having $n_i$ ``successes'' in $N_{\mathsf{all}}$
trials is calculated, where $N_{\mathsf{all}}$ is the total number of
events.  The probability of an individual ``success'' is given by
$F(k_{\mathsf{adc}})$.

The probability distribution for the number of electrons after the
fourth dynode is calculated without approximation. Then the function
is scaled, approximating the additional spread due to the remaining
dynodes with a Gaussian, as described in the previous section.
\subsection{MC--Generated Spectra}
\begin{table*}
{\setlength{\tabcolsep}{0em}
\begin{center}
\parbox{\textwidth}{\caption{%
Voltage distribution in 12-dynode PMT, normalised to the voltage
between dynodes 3 and 4.
\label{tab:mapmt.dynodechain}}}
\begin{tabular}{ |c| *{23}{p{1em}|}}
\hline
\hspace{0.5em}voltage\hspace{0.5em}   
& \multicolumn{1}{p{5em}|}{\mbox{} } &
\multicolumn{2}{p{2em}|}{\hspace{0.7em}3} &
\multicolumn{2}{p{2em}|}{\hspace{0.7em}2} &
\multicolumn{2}{p{2em}|}{\hspace{0.7em}2} &
\multicolumn{2}{p{2em}|}{\hspace{0.7em}1} &
\multicolumn{2}{p{2em}|}{\hspace{0.7em}1} &
\multicolumn{2}{p{2em}|}{\hspace{0.7em}1} &
\multicolumn{3}{p{3em}|}{\hspace{0.5em}$\cdots$} &
\multicolumn{2}{p{2em}|}{\hspace{0.7em}1} &
\multicolumn{2}{p{2em}|}{\hspace{0.7em}1} &
\multicolumn{2}{p{2em}|}{\hspace{0.7em}2} &
\\
\hspace{0.5em}dynode number\hspace{0.7em}
& \multicolumn{2}{p{6em}|}{\hspace{0.5em} Cathode }
& \multicolumn{2}{p{2em}|}{\hspace{0.7em}1}
& \multicolumn{2}{p{2em}|}{\hspace{0.7em}2}
& \multicolumn{2}{p{2em}|}{\hspace{0.7em}3}
& \multicolumn{2}{p{2em}|}{\hspace{0.7em}4}
& \multicolumn{2}{p{2em}|}{\hspace{0.7em}5}
& \multicolumn{5}{p{5em}|}{\hspace{1.5em}$\cdots$}
& \multicolumn{2}{p{2em}|}{\hspace{0.3em}10}
& \multicolumn{2}{p{2em}|}{\hspace{0.3em}11}
& \multicolumn{2}{p{2em}|}{\hspace{0.3em}12}\\
\hline
\end{tabular}
\end{center}
}
\end{table*}
 The validity of the the method has first been established on Monte
 Carlo simulated data.
 The Monte Carlo program simulates the output of a PMT pixel. The
 gain at the first dynode is $g_1=5$ and the gains at the other
 dynodes are calculated from $g\propto V^\alpha$ with $\alpha=0.75$.
 The values for $V$ are given in table \ref{tab:mapmt.dynodechain}.

\begin{figure}
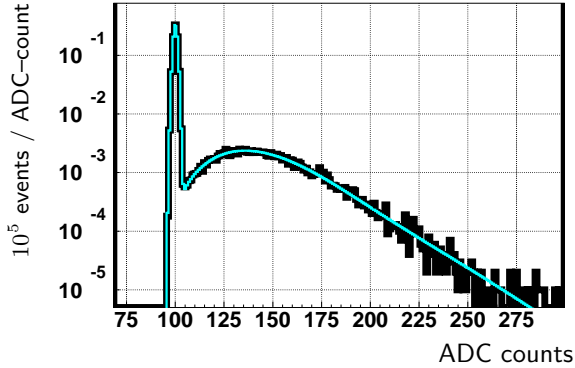

\begin{center}
\caption{%
MC--generated PMT ADC--spectrum, from 100k events, with
$\lambda_1=0.15$. The fit is superimposed.
\label{fig:func_fitmc}}
\labelpaw{%
\fig/fitted_mc_12dyns_gain5}{%
\columnwidth}{%
\columnwidth}{%
\textsf{%
 ADC counts \hspace{0.5em}}}{%
\textsf{%
\small $10^{5}$ events / ADC--count\hspace{0.5em}}}{
-2mm}
\end{center}
\end{figure}

 The fit function is applied to two sets of 128 simulations with $10^5$
 events each, one set with $0.15$ photoelectrons per event, one with
 $3.0$ photoelectrons per event.  A spectrum from each set is shown in
 figures \ref{fig:func_fitmc} and \ref{fig:func_fitmc_l3}.

 The fits are performed varying the gain of only one dynode and
 calculating the gains at the other dynodes using the same value for
 $\alpha$ as in the Monte Carlo program that generated the spectrum.
\begin{table}
\begin{center}
\caption{%
Monte Carlo input compared with mean and RMS of the results from fits to 128
simulated spectra, with $\lambda_1=0.15$
\label{tab:func_mc_fit_comparison}}
\begin{tabular}{ || l |c | r@{$\pm$}l || }
\hline\hline
                     & MC input & \multicolumn{2}{c||}{\parbox{3.5cm}
                     {Mean fit result\\
                     $\pm$ RMS spread}}\\
\hline
$\lambda_1$          & 0.150    & 0.1501  & 0.0013               \\
$g_1$                & 5.000    & 5.0012  & 0.058               \\
$p_0$                & 100.00   & 99.999  & 0.0038              \\
$\sigma$             & 1.0000   & 1.0004  & 0.0027               \\
$c_n$                & 
                     $3.20\cdot 10^{-4}$ & 
                     $(3.23$& $0.46)\cdot 10^{-4} $ \\
\hline\hline
\end{tabular}
\end{center}
\end{table}
\begin{figure}
\begin{center}
\caption{%
MC--generated PMT ADC--spectrum, from 100k events, with
$\lambda_1=3$. The fit is superimposed.
\label{fig:func_fitmc_l3}}
\labelpaw{%
\fig/fitted_mc_12dyns_gain5_lambda3}{%
\columnwidth}{%
\columnwidth}{%
\textsf{%
 ADC counts \hspace{0.5em}}}{%
\textsf{%
\small $10^{5}$ events / ADC--count\hspace{0.5em}}}{
-2mm}
\end{center}
\end{figure}
\begin{table}
\begin{center}
\caption{%
Monte Carlo input compared with mean and RMS of the results from fits
to 128 simulated spectra (representing 2 64--channel MaPMT's), with
$\lambda_1=3$
\label{tab:func_mc_fit_comparison_l3}}
\begin{tabular}{ || l |c | r@{$\pm$}l || }
\hline\hline
                     & MC input & \multicolumn{2}{c||}{\parbox{3.5cm}
                     {Mean fit result\\
                      $\pm$ RMS spread}}\\
\hline
$\lambda_1$          & 3.000    & 3.002   & 0.022               \\
$g_1$                & 5.000    & 4.985   & 0.107               \\
$p_0$                & 100.000  & 99.999  & 0.021               \\
$\sigma$             & 1.000    & 0.999  & 0.016               \\
$c_n$                & 
                     $6.4\cdot 10^{-4}$ & 
                     $(6.45$& $0.17)\cdot 10^{-4} $ \\
\hline\hline
\end{tabular}
\end{center}
\end{table}
The fit results agree very well with the input values, as
shown in tables \ref{tab:func_mc_fit_comparison} and
\ref{tab:func_mc_fit_comparison_l3}.
To test the sensitivity of the fit result on the exact knowledge of
$\alpha$, the fit to the spectrum in figure
\ref{fig:func_fitmc} is repeated assuming different values for this
parameter in the fit--function: $\alpha=0.5$ and $\alpha=1.0$. The
results are given in table \ref{tab:func_mc_fit_comparison_wrongk}.
\begin{table*}
\begin{center}
\caption[Monte Carlo input compared with fit using different assumptions in the
fit.]{Monte Carlo input compared with fit--result for the MC--generated
pulse height spectrum shown in figure \ref{fig:func_fitmc}, using
different assumptions in the fit.
\label{tab:func_mc_fit_comparison_wrongk}}
{\small
\begin{tabular}{ || l |c *{3}{|c} |r@{$\pm$}l || }
\hline\hline
                     & \parbox{2cm}{MC input \\$\alpha=0.75$} & 
\parbox{1.6cm}{Fit result \\$\alpha=0.75$}&
\parbox{1.6cm}{Fit result \\$\alpha=0.5$} &
\parbox{1.6cm}{Fit result \\$\alpha=1$}  &
\multicolumn{2}{|c||}{\parbox{2.8cm}{Fit result: 3 \\ indep. dyn's}}\\
\hline
$\lambda_1$          & 0.1500 &
                       0.1490 & 
                       0.1491 &
                       0.1489 &
                       0.1492 & 0.0013 \\
$g_1$                & 5.00  &
                       5.039 &
                       4.852 & 
                       5.291 & 
                       4.74  & 0.44 \\
\small $g_2,g_3,g_{12}$     
& \small $g_1 \cdot (\frac{2}{3})^{\alpha} =3.69$     & 
                              &
                              &
                              &
                      4.51    & 1.35               \\
\small $g_4,\ldots,g_{11}$  & \small $g_1 \cdot (\frac{1}{3})^{\alpha} =2.19$     & 
                             &
                             &
                             &
                     1.97    & 0.21               \\
$p_0$                & 100.000  &
                       100.000  & 
                       100.000&
                       100.000&
                       100.000& 0.003              \\
$\sigma$             & 1.0000 &
                       1.0028 &
                       1.0029 &
                       1.0027 &
                       1.0028 & 0.0025             \\
$c_n$                & 
                     $ 3.20\cdot 10^{-4}$  &
                     $ 2.90\cdot 10^{-4}$  &
                     $ 0.373\cdot 10^{-4}$ &
                     $ 19.7\cdot 10^{-4}$   &
                     \multicolumn{2}{|c||}{$ (4.37\pm 0.98)\cdot 10^{-4}$} \\
\hline\hline
\end{tabular}
}
\end{center}
\end{table*}
Another fit was performed that does not use the formula $g\propto
V^{\alpha}$. Here it is only assumed that dynodes with the same
accelerating voltage have the same gain. Instead of one gain, three
gains need to be fitted, one for each accelerating voltage. The fits
are performed using the function minimisation and error analysis
package MINUIT \cite{minuit}. The results from this fit, with
error--estimates provided by MINUIT, are given in the last column of
table \ref{tab:func_mc_fit_comparison_wrongk}.

Comparing the results for the different assumptions shows that they
have little impact on the the fitted value for the number of photo
electrons and the gain at the first dynode. Most of the error
introduced by an incorrect estimate of the parameter $\alpha$ is absorbed
into the ratio of ADC--counts to electrons, $c_n$, while the values for
$\lambda_1$ and $g_1$ come out close to the input values.

\subsection{Application to Testbeam Data}

The fit method has been applied to spectra obtained from a prototype
RICH detector, incorporating an array of nine 64--channel Hamamatsu
PMTs and operated in a CERN testbeam \cite{lhcb:rich.tdr}. Fits were
performed to estimate the signal loss at the first dynode and below
the threshold cut.
\begin{figure}
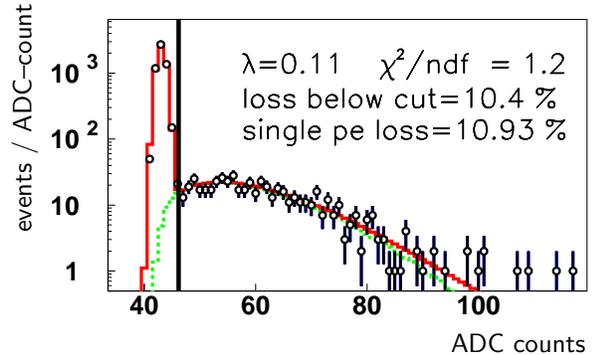

\begin{center}
\caption{%
 Data from 6k events in black, with fit superimposed. The dashed line
 indicates the single photoelectron contribution. The signal loss
 refers to the fraction of photoelectrons lost below the threshold
 cut; both the total fraction of photoelectrons, and the fraction of
 single photoelectron events lost below the cut is given. These
 numbers do not include the loss at the first dynode due to
 photoelectrons producing zero secondary electrons.
\label{fig:func_fitdata}}
\labeltrimmedpaw{\fig/fit_pmt_spectrum_2610_6_32_xfig}{%
\columnwidth}{
0.92\columnwidth}{
\textsf{
 ADC counts\hspace{0.5em}}}{
\textsf{
 events / ADC--count}}{5mm}
\end{center}
\end{figure}
%
%
\begin{table}
\begin{center}
\parbox{\columnwidth}{\caption{%
Result of fit applied to testbeam data \label{tab:func_datafit}}}
\begin{tabular}{ || l | r@{$\pm$}l || }
\hline\hline
                     & \multicolumn{2}{c||}{Fit result}\\
\hline
$\lambda_1$          & 0.107  & 0.005               \\
$g_1$                & 3.60   & 0.20                \\
$p_0$                & 43.06  & 0.01               \\
$\sigma$             & 0.724  & 0.008               \\
$c_n$                & $(61$& $30)\cdot 10^{-4} $ \\
\hline\hline
\end{tabular}
\end{center}
\end{table}
 Figure \ref{fig:func_fitdata} shows an example of such a fit to a
 spectrum obtained in the testbeam. The fit describes the data well,
 with a $\chi^2/\mathsf{dgf}$ of 1.2\footnote{The fit is performed
 with the same log--likelihood method that was used for the MC
 spectra; a $\chi^2$ value is calculated after the fit.}. The line in
 figure \ref{fig:func_fitdata} marks the threshold cut used for photon
 counting in the testbeam. The fraction of single photoelectron events
 below that cut is $\sim 10\%$ (this does not include the
 irrecoverable loss of photoelectrons that do not produce any
 secondaries in the first dynode).
\subsection{Background}
\begin{figure}
\begin{center}
\caption{MC--simulated PMT spectrum with background; the fit
result is superimposed\label{fig:bgmc}}
\labeltrimmedpaw{\fig/dyn_photolectric_withresults}{%
\columnwidth}{
0.92\columnwidth}{
\textsf{
 ADC counts\hspace{0.5em}}}{
\textsf{
 events / ADC--count}}{5mm}
\end{center}
\end{figure}
\begin{figure}
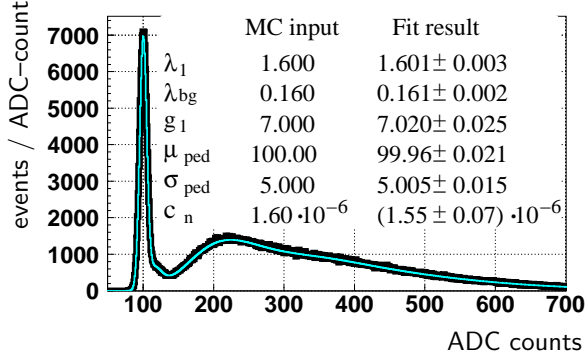

\begin{center}
\caption[The different contributions to the fit]{The different
contributions to the fit in figure
\ref{fig:bgmc}\label{fig:dbmc_decon}}
\labeltrimmedpaw{\fig/dyn_decon_labelled}{%
\columnwidth}{
0.92\columnwidth}{
\textsf{
 ADC counts\hspace{0.5em}}}{
\textsf{
 events / ADC--count}}{5mm}
\end{center}
\end{figure}
 Apart from the Gaussian noise taken into account here, various other
 sources of background, such as electrons released due to the
 photoelectric effect in the first dynode, thermal electrons from the
 photocathode or the dynodes, genuine photoelectrons missing the first
 dynode, etc, can contribute to a PMT pulse height spectrum.  A
 detailed discussion of such background is beyond the scope of this
 paper. However, any type of background that originates from within
 the dynode chain can be naturally accomodated in the fit method
 described here, since this background undergoes the same type of
 amplification process as the signal. To illustrate this, a spectrum
 has been generated with a Monte Carlo program, assuming a signal of
 $1.6$ photoelectrons per event and a background of $0.16$
 photoelectrons per event due to the photoelectric effect in the first
 dynode (see \cite{Chirikov-Zorin:2001qv} for a fit to real data
 showing this kind of background, using a different method).  The
 function to fit this spectrum is obtained by convoluting the
 background--free function $P(k)$ with another function
 $P_{\mathrm{bg}}(k)$. $P_{\mathrm{bg}}(k)$ is identical to $P(k)$
 except that the amplification due to the first dynode is missing and
 that the number of photoelectrons per event hitting the second
 dynode, $\lambda_{\mathrm{bg}}$, is a new free parameter. In the
 example given here, $P$ and $P_{\mathrm{bg}}(k)$ are calculated to
 give the exact distributions for signal and background respectively
 after the fourth dynode; then the two distributions are convoluted
 and the result is scaled according to equation \ref{eqn:scale}. The
 generated spectrum and the fit result are shown in figure
 \ref{fig:bgmc}; the fit function is shown again in figure
 \ref{fig:dbmc_decon} showing the non--pedestal and the single
 photoelectron contributions separately.

\section{Summary}
 An analytical formula for the the probability distribution of the
 number of electrons at the end of a dynode chain, or any ``snowball''
 like process described by a series of Poisson distributions, is
 derived. The formula describes the amplification process at all
 stages exactly, in particular without approximating Poisson
 distributions with Gaussians. It is evaluated as a function of the
 number of photoelectrons coming from the cathode and the gains at
 each dynode.  The initially found formula is adapted to reduce
 numerical problems due to the multiplication of very large numbers
 with very small ones. A numerical recipe is given that implements
 that function.

 It is shown how the function can be used as the core element of an
 approximated, but faster algorithm, that calculates the exact
 distribution for the first few dynodes and then scales the result
 according to the gain at the remaining dynodes, approximating the
 additional spread at those dynodes with a Gaussian. The number of
 dynodes for which the distribution is calculated exactly is not
 limited in principle and can be adjusted according to the precision
 required, and the computing time available.  It is also shown how to
 modify the function to describe ADC--spectra obtained from read--out
 electronics, rather than directly the number of electrons at the end
 of a dynode chain.

 This fast algorithm is then used to fit Monte Carlo generated
 ADC--spectra. In the fit function, the electron distribution after
 the first four out of twelve dynodes is calculated exactly. The fit
 results reproduce the MC--input values well. The dependence of the
 fit result on the assumptions made to reduce the number of
 fit--parameters is investigated. These results show that the fitted
 value for the number of photoelectrons per event is very weakly
 dependent on the different assumptions considered here, and the
 fitted gain on the first dynode also does not depend strongly on
 them. Real data from a multi--anode PMT used in the 1999 LHCb--RICH
 testbeam are fitted, and shown to be described well by the
 function. Finally it is illustrated how the fit function can be
 modified further to accommodate background from within the dynode
 chain, using the example of the photoelectric effect in the first
 dynode.

\section*{Acknowledgements}
 I wish to thank the LHCb RICH group, and in particular the colleagues
 involved in the 1999 LHCb--RICH testbeam. Special thanks go to James
 Libby, David Websdale and Guy Wilkinson for many helpful suggestions.

\appendix
\section{FORTRAN Routine to Calculate $P(k_n)$}
\label{app:listing}
{\tiny
*\begin{verbatim}
      SUBROUTINE DYNODE_CHAIN(OUT, MAX, LAMBDA, DYNODES)
      IMPLICIT NONE
*     This program takes as its input the maximum number of electrons at
*     the end of the dynode chain, for which it should calculate P(k_n),
*     MAX, the average number of photo-electrons hitting the first
*     dynode, LAMBDA(1), the gains at each dynode, LAMBDA(2),
*     ... LAMBDA(DYNODES) and the dimension of the array LAMBDA:
*     DYNODES. It calls the routine MAKE_P_RATIO, which is listed at the
*     end of this file.
*
*     The output is put into the array OUT(MAX), where the probability
*     to find k_n < MAX electrons at the end of the dynode chain is
*     given by OUT(k_n).
*
*     Written by Jonas Rademacker.
*
      INTEGER MAX, DYNODES
      DOUBLE PRECISION OUT(0:MAX), LAMBDA(DYNODES)
      
      INTEGER ABS_MAX, MAX_DYN
      PARAMETER(ABS_MAX=50001,MAX_DYN=13)

      INTEGER IX,IY,M,I, K, J

*     To avoid having to define a limit on the number k_n that can be
*     calculated, one could create these arrays outside the program and
*     pass them on.
      DOUBLE PRECISION F(1:MAX_DYN) ! corrsponds to f^{star} in the text
      DOUBLE PRECISION U(1:MAX_DYN,0:ABS_MAX),V(1:MAX_DYN,0:ABS_MAX)
      DOUBLE PRECISION X(1:MAX_DYN)

      DOUBLE PRECISION FASTNULL
      PARAMETER (FASTNULL=1.d-300)

      DOUBLE PRECISION MEAN
      
      DOUBLE PRECISION P_ratio(ABS_MAX), F_FACTOR, U_FACTOR, V_FACTOR

      INTEGER MAX_OLD
      SAVE MAX_OLD
      DATA MAX_OLD/-9999/

      SAVE P_ratio

*     -- Some initialisations and tests --
      DO IX=1,MIN(ABS_MAX,MAX),+1
         OUT(IX)=0.d0
      ENDDO
      IF(ABS_MAX.LT.MAX)THEN
         RETURN
      ENDIF
      MEAN = 1.D0
      DO IX=1,DYNODES,+1
         MEAN = MEAN*LAMBDA(IX)
      ENDDO
      IF(MEAN.LE.0.d0)THEN
         OUT(0)=1.d0
         RETURN
      ENDIF

*     -- make and save the factors P_ratio(k)=(p_{k}/p_{k-1})^{k} --
      IF(MAX.GT.MAX_OLD)THEN
         MAX_OLD=MAX
         CALL MAKE_P_RATIO(P_ratio,MAX)
      ENDIF

*     -- Calculate the probability to see zero electrons (k_n=0) --
      F(DYNODES)=1.d0
      U(DYNODES,0)=F(DYNODES)
      V(DYNODES,0)=F(DYNODES)
      DO IX=DYNODES-1,1,-1
         X(IX)   = LAMBDA(IX)*DEXP(-LAMBDA(IX+1))
         F(IX)   = DEXP(X(IX)*F(IX+1))
         U(IX,0) = F(IX)
         V(IX,0) = F(IX)
      ENDDO
      OUT(0)=DEXP(-LAMBDA(1))*F(1) !   <---- save the result

*     -- Calculate the probabilities for k_n=1,...,MAX electrons --
      DO K=1,MAX,+1
*     .  calculate f_n
         IF(F(DYNODES).LT.FASTNULL)THEN
            F(DYNODES)=0.d0
         ELSE
            F(DYNODES)=F(DYNODES) * LAMBDA(DYNODES)/DBLE(K)
         ENDIF
         U(DYNODES,K)=F(DYNODES)
         V(DYNODES,K)=F(DYNODES)

*     .  re-calculate U and V from previous iteration:
         DO J=0,K-1,+1
            F_FACTOR=P_ratio(K)**(DBLE(J)/DBLE(K))
            IF(K-1-J.GT.0)THEN
               U_FACTOR=DSQRT(DBLE(K-1)/DBLE(K-1-J))*
     &              F_FACTOR
            ELSE
               U_FACTOR=F_FACTOR
            ENDIF
            V_FACTOR=DSQRT((DBLE(K-1)/DBLE(K-J)))*
     &           F_FACTOR

            DO I=DYNODES,1,-1
               U(I,J)=U(I,J)*U_FACTOR
               V(I,J)=V(I,J)*V_FACTOR
            ENDDO
         ENDDO

*     .  apply the recursive formula to get f^{k}_i
         DO I=DYNODES-1, 1, -1
            F(I)=0.d0
            DO J=0,K-1
               F(I)=F(I)+U(I,K-1-J)*X(I)*V(I+1,J+1)
            ENDDO
            U(I,K)=F(I)
            V(I,K)=F(I)
         ENDDO
*     .  calculate P(k):
         OUT(K)=DEXP(-LAMBDA(1))*F(1) !   <---- save the result
      ENDDO

      RETURN
      END

*__________________________________________________________________
      SUBROUTINE MAKE_P_RATIO(P_ratio,MAX)
      IMPLICIT NONE
      INTEGER MAX
      DOUBLE PRECISION P_ratio(MAX)

      INTEGER N
      DOUBLE PRECISION NFAC

      DOUBLE PRECISION PI, E
      PARAMETER(PI=3.1415927d0, E=2.718281828d0)

      INTEGER APPROX_FROM
      PARAMETER(APPROX_FROM=25)

      NFAC=1.D0
      P_ratio(1)=1.d0
      DO N=2,MIN(APPROX_FROM-1,MAX),+1
         NFAC=NFAC*DBLE(N-1)
         P_ratio(N)=(NFAC**(1.d0/DBLE(N-1)))/DBLE(N)
      ENDDO

      DO N=APPROX_FROM,MAX,+1
         P_ratio(N)=
     &        (2.D0*PI*DBLE(N-1))**(1.D0/(2.D0*DBLE(N-1)))*
     &        DBLE(N-1)/(E*DBLE(N))*
     &        (1.d0+1.d0/DBLE(12*(N-1))+
     &        1.d0/DBLE(288*(N-1)**2)
     &        )**(1.D0/DBLE(N-1))
      ENDDO

      RETURN
      END

*________________________________________________________________
*\end{verbatim}

}

\bibliographystyle{phys}
\bibliography{bibliography}

\newcommand{\etalchar}[1]{$^{#1}$}
\begin{thebibliography}{CZ{\etalchar{+}}01}

\bibitem[CZ{\etalchar{+}}01]{Chirikov-Zorin:2001qv}
I.~Chirikov-Zorin et~al.
\newblock Method for precise analysis of the metal package photomultiplier
  single photoelectron spectra.
\newblock {\em Nucl. Instrum. Meth.}, {\bf A456}:310, 2001.

\bibitem[Ham00]{hamamatsu}
{\em Hamamatsu Book on Photo Multipliers}, September 2000.

\bibitem[Jam94]{minuit}
F.~James.
\newblock {\em MINUIT Function Minimization and Error Analysis. Reference
  Manual. Version 94.1}, March 1994.
\newblock CERN Program Library Long Writeup D506.

\bibitem[RIC00]{lhcb:rich.tdr}
{\em {L}{H}{C}{b} RICH, Technical Design Report}, September 2000.
\newblock CERN/LHCC/2000-0037.

\end{thebibliography}

\end{document}